\documentclass[epj]{svjour}
\newcommand{\PD}{\partial}
\newcommand{\be}{\begin{equation}}
\newcommand{\ee}{\end{equation}}
\newcommand{\bea}{\begin{eqnarray}}
\newcommand{\eea}{\end{eqnarray}}

\usepackage{graphics}
\usepackage{epsfig}
\usepackage{epsf}

\begin{document}
\title{Dynamical Correlation Length near the Chiral Critical Point}
\author{Kerstin Paech}
%
%\offprints{}          % Insert a name or remove this line
%
\institute{Institut f\"ur Theoretische Physik, J.W.~Goethe Universit\"at,
Postfach 111932, D-60054 Frankfurt am Main, Germany}
\date{Received: date / Revised version: date}
\abstract{
The dynamical evolution of small systems undergoing a
chiral symmetry breaking transition in the course of rapid
expansion is discussed. The time evolution of the dynamical
correlation length for trajectories passing
through a second-order critical point is extracted.
It is shown that while the maximum value of the
correlation length is bound from above by dynamical effects,
the time interval during which it is near its maximum grows
steadily with the system size and with decreasing expansion
rate.
}
\maketitle
\section{Introduction}
\label{intro}
The main goal of colliding heavy ions at high energies is to produce
matter at high temperature and baryon density. In matter at such extreme
conditions chiral symmetry may be approximately restored (for a review
of signatures for hot and dense matter see e.g.~\cite{Scherer:1999qq}). 
In particular,
it has been argued that a line of first-order phase transitions in the
phase diagram of QCD could end in a second-order critical point, where the
correlation length diverges~\cite{Stephanov:1998dy}. The critical 
point in the temperature
vs. baryon-chemical potential plane has been located on the
lattice~\cite{Fodor:2001pe}. Here, we study the
dynamics of the chiral fields near a critical point and determine the
behavior of the correlation length in finite and rapidly expanding systems,
such as the ones encountered in heavy-ion collisions.

\section{Chiral Hydrodynamics}
\label{model}
In Chiral Hydrodynamics~\cite{Mishustin:1997iz,Paech:2003fe} it is assumed
that the long-wavelength (classical) modes of the chiral fields evolve
in the effective potential generated by the thermalized degrees of freedom,
which are the matter fields (and possibly also hard modes of the chiral field).
In our model, the latter are described as a perfect
relativistic fluid, whose equation of state is in turn determined by the
chiral field (via the effective mass), and which can exchange
energy and momentum with the chiral fields.
The chiral symmetry breaking dynamics is described by an effective
field theory, in our case the SU(2)$\times$SU(2) linear $\sigma$-model:
\bea
{\cal L} &=&
 \overline{q}\,\left[i\gamma ^{\mu}\partial _{\mu}-g(\sigma +\gamma _{5}
 \vec{\tau} \cdot \vec{\pi} )\right]\, q\nonumber\\
&+& \frac{1}{2}\left (\partial _{\mu}\sigma \partial ^{\mu}\sigma + 
\partial _{\mu} \vec{\pi} \partial ^{\mu}\vec{\pi} \right)
- U(\sigma ,\vec{\pi})\quad.
\label{sigma}
\eea
The potential $U(\sigma ,\vec{\pi})$, which exhibits both spontaneously and
explicitly broken symmetry, is given by
\begin{equation} \label{T=0_potential}
U(\sigma ,\vec{\pi} )=\frac{\lambda ^{2}}{4}(\sigma ^{2}+\vec{\pi} ^{2} -
{\it v}^{2})^{2}-h_q\sigma\quad.
\end{equation}
Here $q$ is the constituent-quark field $q=(u,d)$. The
scalar field $\sigma$ and the pseudoscalar field $\vec{\pi}$ 
together form a chiral field $\phi = (\sigma,\vec{\pi})$. 
The vacuum expectation values of the condensates are
$\langle\sigma\rangle
={\it f}_{\pi}$ and $\langle\vec{\pi}\rangle =0$, where ${\it f}_{\pi}=93$~MeV 
is the pion decay constant. 

For $g>0$, the finite-temperature one-loop effective potential
includes a contribution from the quarks, and is given by
\be
V_{\rm eff}(\phi,T) =
U(\phi) - d_{q} \int \! \frac{d^3{\vec{p}}}{(2\pi)^3}
T \log \left[1+e^{{-E}/{T}} \right]\nonumber
\label{effpot}
\ee
$V_{\rm eff}$ depends on the order parameter field through
the effective mass of the quarks, $m_q=g|\phi|$, which enters
the expression for the energy $E$.

For sufficiently small $g$ one finds a smooth
transition between the two phases. For large coupling, however,
the effective potential exhibits a first-order phase
transition~\cite{Scavenius:1999zc}.
Along the line of first-order transitions the effective potential has
two minima. At $T=T_c$ these two minima are degenerate and are seperated by
a barrier. As one lowers the value of $g$ the barrier gets smaller and the
two minima approach each other. At $g_c\simeq 3.7$,
finally, the barrier vanishes, and so does the latent heat.
Below, we study the hydrodynamic expansion near this chiral
critical point.

The classical equations of motion for the chiral
fields are
%\bea
\be
\partial_{\mu}\partial^{\mu}\phi+ \frac{\delta V_{\rm eff}}{\delta\phi}=0\nonumber\\
%\partial_{\mu}\partial^{\mu}\sigma+ \frac{\delta V_{\rm eff}}{\delta\sigma}=0\nonumber\\
%\partial_{\mu}\partial^{\mu}\vec{\pi}+\frac{\delta V_{\rm eff}}{\delta\vec\pi}=0
\label{EulerLagrange}
\ee
%\eea
The dynamical evolution of the thermalized degrees of freedom (fluid of 
quarks) is
determined by the local conservation laws for energy and momentum.
Note that we do not assume that the chiral fields are in equilibrium
with the heat bath of quarks. Hence, the fluid pressure $p$ depends explicitly
on $|\phi|$, see~\cite{Paech:2003fe} for more details. 
Due to the interaction between fluid and field the {\em total} energy and 
momentum is the conserved quantity:
\be \label{contEq_Q}
\PD_\mu \left( T^{\mu\nu}_{\rm fluid} + T^{\mu\nu}_\phi \right)=
0.
\ee
We emphasize that we employ eq.~(\ref{EulerLagrange}) not only
to propagate the mean field through the transition but fluctuations as
well. The initial condition includes some generic ``primordial''
spectrum of fluctuations (see below) which then
evolve in the effective potential generated by the matter fields.
Near the critical point, those fluctuations
have small effective mass and ``spread out'' to probe the flat
effective potential.

\section{Correlation Length}

The correlation length provides a measure for the typical length scale 
over which fluctuations of the fields are correlated.
Following the Ginzburg-Landau theory, for an infinite system
in global thermal equilibrium
one can expand the free energy $F$ arround the thermodynamic 
expectation value $\phi_0$ - which is given by the minimum of 
the free energy - in a power series:
\be
F (\bar{\phi})
= a_0 + a_1\bar{\phi} + a_2\bar{\phi}^2  + a_3\bar{\phi}^3  + 
\label{expandf}
a_4\bar{\phi}^4 + \dots + a_{\rm N}\bar{\phi}^N
\ee
where $\bar{\phi}=\frac{\phi-\phi_0}{f_\pi}$ and 
$a_{\rm N}>0$. 
The correlation length $\xi$ is defined as the second derivative of
the free energy
\be
\frac{1}{\xi^2}=\left. \frac{d^2 F(\bar{\phi})}{d\bar{\phi}^2}
\right|_{\bar{\phi}=0} = 2a_2.
\ee
The correlation length is finite
for $T\neq T_c$ since the effective potential has a finite curvature
there. At $T_c$ the potential about $\phi_0$ becomes flat
and thus the correlation length diverges as the system
approaches the critical point. Of course,  this can only be
true for an infinite system in global thermodynamic equilibrium.
In high-energy heavy ion collisions instead one deals with finite and
rapidly expanding systems where the true dynamical
correlation length is finite.

\section{Results}
\subsection{Initial Conditions}
\label{sec_IniCond}
To study the effects of a finite and expanding system we choose
the following initial conditions.
The distribution of the fluid energy density at $t=0$ is taken to be
spherical:
\be
e(t=0,r) = e_{\rm eq} \,\Theta(R-r)~,
\ee
where $e_{\rm eq}$ denotes the equilibrium value of the energy density
taken at a temperature of $T_i\approx 160$~MeV which is well above
the transition temperature $T_c \approx 138$~MeV.
$R$ denotes the initial size of the system.
For the fluid velocity we assume a linear profile
\be
v_r(t=0,r) = \frac{r}{R}\,\Theta(R-r)~.
\ee
Note that the expansion rate (Hubble constant) equals $1/R$, hence larger
systems also correspond to less rapid expansion.

The initial conditions for the chiral fields are 
\bea
\sigma(t=0,\vec{x}) &=& \delta\sigma(\vec{x}) + f_\pi + \nonumber\\
& & {}(- f_\pi + \sigma_{\rm eq})
\cdot \left[ 1 +
\exp\left(\frac{r-R}{a}\right)
\right]^{-1} \nonumber\\
\vec\pi(t=0,\vec{x}) &=& \delta\vec{\pi}(\vec{x})\, , \label{ini_field}
\eea
$a=0.3$~fm is the surface thickness of this Woods-Saxon like distribution.
Here $\sigma_{\rm eq}\approx0$ is the value of the $\sigma$ field
corresponding to $e_{\rm eq}$.
Thus, the chiral condensate nearly vanishes at the center, where the energy
density of the quarks is large, and then quickly interpolates to $f_\pi$
where the matter density is low.

$\delta\sigma(\vec{x})$ and $\delta\vec{\pi}(\vec{x})$ represent
Gaussian random fluctuations of the fields.
We correlate them over
an initial correlation length of $\xi_0\approx1.2$~fm as described
in~\cite{Paech:2003fe}. Our focus now is on how
those ``primordial'' fluctuations evolve through the transition
and how the correlation length depends on the system size.

\subsection{Extraction of Correlation Length}
To analyse the time evolution of $\xi$ in our finite
and expanding system we need to know the {\it dynamical} effective potential
probed by the fluctuating fields. This is done by extracting a histogram of
the field distribution at every time step, within a sphere
of radius 1~fm around $r=0$. This histogram has been averaged over
a few random initial field configurations, picked according to
eq.~(\ref{ini_field}).

The probability distribution is related to the 4-d effective action
by 
\be
P[\phi]\propto\exp\left\{-S_{\rm eff}\right\}~,
\ee
and thus we can extract the effective action ''seen'' by the chiral
fields by fitting a polynomial of the form~(\ref{expandf}) to
\be
S_{\rm eff}(\phi) \propto - \log\{P[\phi]\}\quad .
\ee

\subsection{Numerical Results}

\begin{figure}
\resizebox{0.5\textwidth}{!}{%
\includegraphics{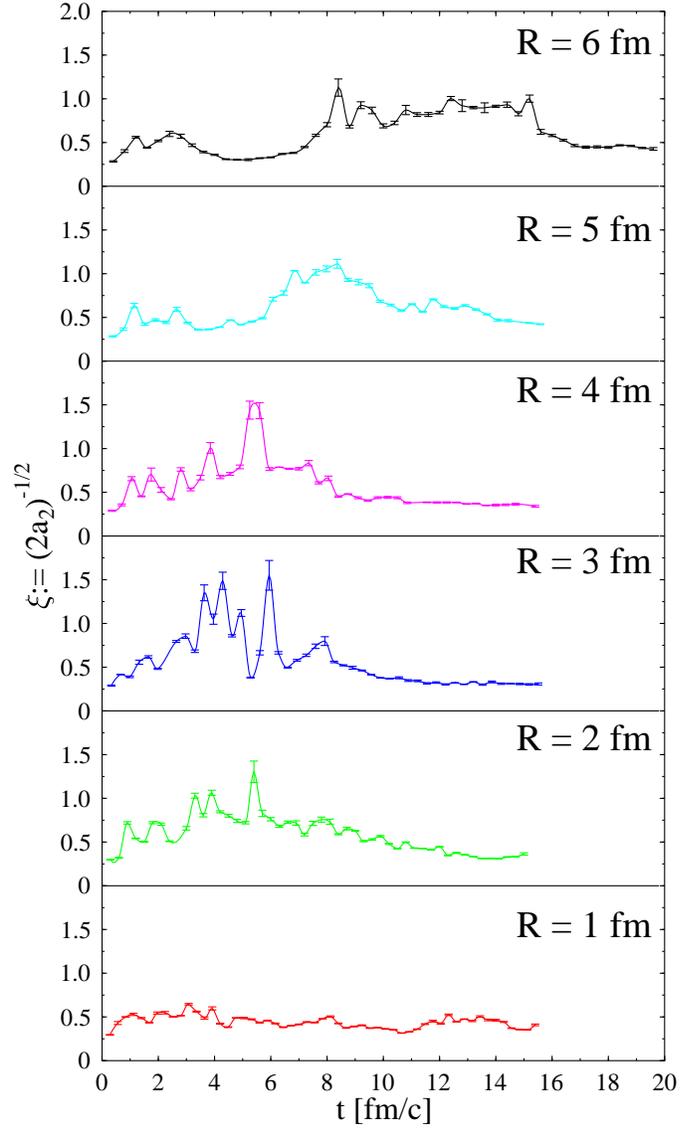}
}
\caption{Time evolution of the correlation length $\xi$ for
finite and expanding systems of different initial radius $R$.}
\label{correlation}
\end{figure}

Fig.~\ref{correlation} depicts the time evolution of the extracted
correlation length $\xi$ for different system sizes.
Evidently, the correlation length is always {\em finite}.
For a system of initial radius $R=1$~fm, $\xi(t)\approx\xi_0$
remains approximately constant during the expansion, equal to the initial
correlation length. 
However, for larger $R$, it develops a maximum at intermediate
times, roughly twice $\xi_0$. Its maximum value appears to grow
only very slowly with $R$ (once $R\ge 3$~fm), i.e.\ the approach towards
$\xi=\infty$ for $R\to\infty$ is slow. Both observations are in line 
with~\cite{Berdnikov:1999ph} who studied finite-time effects for infinitely
large systems, and estimated that $\xi$ can not exceed $2-3$ times
$\xi_0$, and that the maximum value of $\xi$ grows only slowly with decreasing
cooling rate. This implies that experimental observables will not exhibit
``fully developped'' critical behavior even for trajectories through the
critical point. On the other hand, one might hope for some signals to show up
even for trajectories that miss the critical point somewhat, since the growth
of $\xi$ is damped by dynamical effects anyways. We also observe from our
real-time analysis that the time interval during which $\xi$ is near its
maximum {\it grows} steadily with $R$. This might be relevant for observables
that integrate over the entire collision history.  

The behavior of other couplings $a_i$ ($i>2$) belonging to marginal
or irrelevant operators will be reported elsewhere.

\section*{Acknowledgment}
I thank the organizers for support and the opportunity to attend EPS-2003,
Adrian Dumitru for continuous motivation and advice, 
the Josef Buchmann Foundation for a fellowship,
and the Frankfurt Center 
for Scientific Computing for providing computational resources.
% BibTeX users please use

\providecommand{\href}[2]{#2}\begingroup\raggedright\endgroup
\end{document}